# Query Data With Fuzzy Information In Object-Oriented Databases An Approach The Semantic Neighborhood Of Hedge Algebras

Doan Van Thang

Korea-VietNam Friendship Information Technology College
Department of Information systems, Faculty of Computer
Science Da Nang City, Viet Nam
vanthangdn@gmail.com

Doan Van Ban

Institute of Information Technology, Academy Science and
Technology of Viet Nam.
Ha Noi City, Viet Nam

*Abstract -* **In this paper, we present an approach for handling attribute values of object classes with fuzzy information and uncertainty in object-oriented database based on theory hedge algebraic. In this approach, semantics be quantified by quantitative semantic mapping of hedge algebraic that still preserving in order semantics may allow manipulation data on the real domain of attribute in relation with the semantics of linguistic. And then, evaluating semantics, searching information uncertainty, fuzziness and classical data entirely consistent based on the ensuring homogeneity of data types. Hence, we present algorithm that allow the data matching helping the requirements of the query data.**

## I. Introduction

In approach interval value [2], we consider to attributive values object class is interval values and the interval values are converted into sub interval in [0, 1] respectively and then we perform matching interval this. However, attributive value of the object in the fuzzy object-oriented database is complex: linguistic values, reference to objects (this object may be fuzzy), collections,… Thus matching data also become more complex. Hence, query information method proposed in [2] is not satisfy requirements for the case of this data yet.

In this paper, we research has expanded for handling attribute value is linguistic value. There are many approaches on handling fuzzy information with linguistic sematic that researchers interests [1], [3]. We based on approach hedge algebra, where linguistic semantic is obtained by considering the terms as expressed by the part of order relation. In this approach linguistic value is data which is not label of fuzzy set representation sematic of linguistic value. Using quantitative semantics mapping of hedge algebra to transfer linguistic values into real values that preserve in order semantics may allow manipulation data on the real domain of attribute in relation with the semantics of linguistic.

The paper is organized as follows: Section 2 presenting the basic concepts relevant to hedge algebraic as the basis for the next sections; section 3 proposing **SASN** (**S**earch **A**ttributes in the **S**emantic **N**eighborhood) and **SMSN** (**S**earch **M**ethod in the **S**emantic **N**eighborhood) algorithms for searching data fuzzy conditions for both attributes and methods; section 4 presenting examples for searching data with fuzzy information, and finally conclusion.

## II. Fundamental concepts

In this section, we present some fundamental concepts related to hedge algebra [5].

Let hedge algebra $\underline{X} = ( X, G, H, \leq )$, where $X = LDom(\underline{X})$, $G = \{1, c^{-}, W, c^{+}, 0\}$ is set generator terms, $H$ is a set of hedge considered as a one-argument operations and $\leq$ relation on terms (fuzzy concepts) is a relation order "induced" from natural semantics on $X$. Set $X$ is generated from $G$ by means of one-argument operations in $H$. Thus, a term of $X$ represented as $x = h_n h_{n-1}.......h_1 x$, $x \in G$. Set of terms is generated from the an $X$ term denoted by H$(x)$. Let set hedges $H = H^{-} \cup H^{+}$, where $H^{+} = \{h_1,..., h_p\}$ and $H^{-} = \{h_{-1}, ..., h_{-q}\}$ are linearly ordered, with $h_1 < .. < h_p$ and $h_{-1} < .. < h_{-q}$, where $p, q > 1$, we have the following definitions related:

**Definition 2.1** An $fm : X \rightarrow [0,1]$ is said to be a fuzziness measure of terms in X if:

(1) fm is called complete, that is $\forall u \in X$,

$$\sum_{-q \leq i \leq p, i \neq 0} fm(h_i u) = fm(u)$$

(2) if x is precise, that is H(x) = {x} then $fm(x) = 0$. Hence $fm(0)=fm(W)=fm(1)=0$.

(3) $\forall x, y \in X$, $\forall h \in H$, $\dfrac{fm(hx)}{fm(x)} = \dfrac{fm(hy)}{fm(y)}$, *This*

*proportion is called the fuzziness measure of the hedge h and denoted by $\mu(h)$.*

**Definition 2.2** (Quantitative semantics function $v$)

Let $fm$ is fuzziness measure of $X$, quantitative semantics function $v$ on $X$ is defined as follows:

(1) $v(W) = \theta = fm(c^{-})$, $v(c^{-}) = \theta - \alpha fm(c^{-})$ and $v(c^{+}) = \theta + \alpha fm(c^{+})$

(2) If $1 \leq j \leq p$ then:

$$v(h_j x) = v(x) + Sign(h_j x) \times \left[ \sum_{i=1}^{j} fm(h_i x) - \omega(h_j x) fm(h_j x) \right]$$

(3) If $-q \leq j \leq -1$ then:

$$v(h_j x) = v(x) + Sign(h_j x) \times \left[ \sum_{i=j}^{-1} fm(h_i x) - \omega(h_j x) fm(h_j x) \right]$$

Where:

$$\omega(h_j x) = \frac{1}{2}\Big[1 + Sign(h_j x)Sign(h_\alpha h_j x)(\beta - \alpha)\Big] \in \{\alpha, \beta\}$$

**Definition 2.3** Invoke *fm* is fuzziness measure of hedge algebra X, *f*: X -> [0, 1]. ∀x ∈ X, denoted by $I(x) \subseteq [0, 1]$ and |I(x)| is measure length of I(x).

A family J = {I(x):x∈ X} called the partition of [0, 1] if:

(1): {I(c⁺), I(c⁻)} is partition of [0, 1] so that |I(c)| = *fm*(c), where c∈ {c⁺, c⁻}.

(2): If I(x) defined and |I(x)| = fm(x) then {I(hix): I = 1...p+q}is defined as a partition of I(x) so that satisfy conditions: |I(h_i x)| = *fm*(h_i x) and |I(h_i x)| is linear ordering.

Set {I(h_i x)} called the partition associated with the terms x. We have

$$\sum_{i=1}^{p+q} |I(h_i x)| = |I(x)| = fm(x)$$

**Definition 2.4** Set $X_k = \{x \in X : |x| = k\}$, consider $P^k = \{I(x) : x \in X_k\}$ is a partition of [0, 1]. Its said that u equal v at k level, denoted by u =_k v, if and only if $I(u)$ and $I(v)$ together included in fuzzy interval k level. Denote ∀u, v ∈ X, $u =_k v \Leftrightarrow \exists \Delta^k \in P^k : I(u) \subseteq \Delta^k$ and $I(v) \subseteq \Delta^k$.

## III. DATA SEARCH METHOD

Let fuzzy class C = ({a₁, a₂, ..., aₙ}, {M₁, M₂, ..., Mₘ}); *o* is object of fuzzy class C. Denoted o.aᵢ is attribute value of *o* on attribute aᵢ (1 ≤ i ≤ n) and o.Mⱼ is value method of *o* (1 ≤ j ≤ m).

In [2] we presented the attribute values are 4 cases: precise value; imprecise value (or fuzzy); object; collection. In this paper, we only interested in handing case 1 and 2: precise value and imprecise value (fuzzy value) and to see precise value is particular case of fuzzy value. Fuzzy value is complex and linguistic label is often used to represent the value of this type. Domain fuzzy attribute value is the union two components:

$Dom(a_i) = CDom(a_i) \cup FDom(a_i)$ (1 ≤ i ≤ n).

Where:

- $CDom(a_i)$: domain crisp values of attribute aᵢ.

- $FDom(a_i)$: domain fuzzy values of attribute aᵢ.

### A. Neighborhood level k

We can get fuzzy interval of terms length k as the similarity between terms. It means that the term that representative value of them depending on fuzzy interval *level k* is similar *level k*. However, to build the fuzzy interval *level k*, representative value of terms *x* have length less than k is always in the end of fuzzy interval *level k*. Hence, when determining *neighborhood level k*, we expect representative value it must be inner point of *neighborhood level k*.

Based on fuzzy interval level *k* and *k+1* we construct a partition of the domain [0, 1] following as [8]:

(1) *Similar level 1*: with k = 1, fuzzy interval *level 1* including I(c⁻) and I(c⁺). fuzzy interval *level 2* on interval I(c⁺) is $I(h_{-q}c^+) \le I(h_{-q+1}c^+) ... \le I(h_{-2}c^+) \le I(h_{-1}c^+) \le \upsilon_A(c^+) \le I(h_1 c^+) \le I(h_2 c^+) \le ... \le I(h_{p-1}c^+) \le I(h_p c^+)$. Meanwhile, we construct partition at *similar level 1* include the equivalence classes following: $S(\boldsymbol{0}) = I(h_p c^-)$; $S(c^-)=I(c^-) \setminus [I(h_{-q}c^-) \cup I(h_p c^-)]$; $S(\boldsymbol{W}) = I(h_{-q}c^-) \cup I(h_{-q}c^+)$; $S(c^+) = I(c^+) \setminus [I(h_{-q}c^+) \cup I(h_p c^+)]$ and $S(\boldsymbol{1}) = I(h_p c^+)$.

We see that except the two end points $\upsilon_A(\boldsymbol{0}) = 0$ and $\upsilon_A(\boldsymbol{1}) = 1$, representative values $\upsilon_A(c^-)$, $\upsilon_A(W)$ and $\upsilon_A(c^+)$ are inner point corresponding of classes *similar level 1* $S(c^-)$, $S(W)$ and $S(c^+)$.

(2) *Similar level 2*: with k = 2, fuzzy interval *level 2* including $I(h_i c^+)$ and $I(h_i c^-)$ with -q ≤ i ≤ p. We have equivalence classes following: $S(\boldsymbol{0}) = I(h_p h_p c^-)$; $S(h_i c^-) = I(h_i c^-) \setminus [I(h_{-q} h_i c^-) \cup I(h_p h_i c^-)]$; $S(\boldsymbol{W}) = I(h_{-q} h_{-q} c^-) \cup I(h_{-q} h_{-q} c^+)$; $S(h_i c^+) = I(h_i c^+) \setminus [I(h_{-q} h_i c^+) \cup I(h_p h_i c^+)]$ and $S(\boldsymbol{1}) = I(h_p h_p c^+)$, with -q ≤ i ≤ p.

By the same, we can construct partition equivalence classes *level k* at any. However, in fact, k ≤ 4 and it means that there is maximum 4 hedges consecutive action onto primary terms c⁻ and c⁺. Precise and fuzzy values will be at the *similar level k* if the representative value of their in the same class *similar level k*.

Hence, *neighborhood level k of fuzzy concept* is determining following: Assuming partition the class *similar level k* is intervals $S(x_1)$, $S(x_2)$, ..., $S(x_m)$. Meanwhile, every fuzzy value *fu* is only and only belong to a similar class. Instance for $S(x_i)$ and called *neighborhood level k* of *fu* and denoted by $FRN_k(fu)$.

### B. Relation matching on domain of fuzzy attribute value

Based on the concept neighborhood, we give the definition of the relation matching between terms in the domain of the fuzzy attribute value.

**Definition 3.1**

Let fuzzy class C determine on the set of attributes A and methods M, aᵢ ⊆ A. o₁, o₂ ∈ C. We say that $o_1.a_i =_k o_2.a_i$ and equal *level k* if:

(1) If $o_1.a_i, o_2.a_i \in CDom(a_i)$ then $o_1.a_i = o_2.a_i$ or existence $FRN_k(x)$ such that $o_1.a_i, o_2.a_i \in FRN_k(x)$.

(2) If $o_1.a_i$ or $o_2.a_i \in FDom(a_i)$, instance for $o_1.a_i$ then we have to $o_2.a_i \in FRN_k(o_1.a_i)$.

(3) If $o_1.a_i, o_2.a_i \in FDom(a_i)$ then $FRN_k(o_2.a_i) = FRN_k(o_1.a_i)$.

**Definition 3.2**

Let fuzzy class C determine on the set of attributes A and methods M, aᵢ ⊆ A. o₁, o₂ ∈ C. We say that $o_1.a_i \ge_k o_2.a_i$ if:

(1) If $o_1.a_i, o_2.a_i \in CDom(a_i)$ then $o_1.a_i \ge o_2.a_i$.

(2) If $o_1.a_i$ and $o_2.a_i \in FDom(a_i)$ then we have to

$o_1.a_i \geq FRN_k(o_2.a_i)$ .

(3) If $o_1.a_i , o_2.a_i \in FDom(a_i)$ then
$FRN_k(o_1.a_i) \geq FRN_k(o_2.a_i)$

### C. Algorithm search data approach to semantic neighborhood

In [2] we presented the structure of fuzzy OQL queries are considered as: *select* <attributes>/<methods> *from* <class> *where* <fc>, where <fc> are fuzzy conditions or combination of fuzzy condition that allow using of disjunction or conjunction operations.

In this paper, we use approaching to semantic neighborhood for determining the truth value of the <fc> and associated truth values.

Example, we consider query following "*show all students are possibly young age*". To answer this query, we perform following:

+ *Step 1*: We construct intervals *similar level k*, $k \leq 4$ because it's a maximum 4 hedges consecutive action onto primary terms $c^-$ and $c^+$.

+ *Step 2*: Determine *neighborhood level k* of fuzzy condition. In the above query, fuzzy condition is possibly young should *neighborhood level 2* of possibly young is *FRN₂*(possibly young*)*, and determine *neighborhood level 2* of fuzzy attribute value is *FRNA₂(attr)*. At last based on definition 3.1, we perform data matching two *neighborhood level 2* of *FRNA₂(attr)* and *FRN₂*(possibly young*)*.

Without loss of generality, we consider on cases multiple fuzzy conditions with notation follow as:

- $\vartheta$ is *AND or OR* operation.

- *fzvalue$_i$* is fuzzy values of the i attribute.

On that basis, we built the **SASN** algorithms

**SASN algorithm:** *search data in cases multiple fuzzy conditions for attribute with operation $\vartheta$* .

**Input**: A class C = ({a$_1$, a$_2$, …, a$_n$}, {M$_1$, M$_2$, …, M$_m$}), C = { o$_1$, o$_2$,…, o$_n$}.

where a$_i$, i = 1…p is attribute, M$_j$ is methods.

**Output**: Set of objects o $\in$ C satisfy condition $\underset{i=1}{\overset{p}{\vartheta}}$ (o.a$_i$= *fzvalue$_i$*).

**Method**
// Initialization.

(1) **For** i = 1 **to** p **do**
(2) **Begin**
(3)     Set $G_{a_i} = \{0, \quad c_{a_i}^-, \quad W, \quad c_{a_i}^+, \quad I\}$;
    $H_{a_i} = H_{a_i}^+ \cup H_{a_i}^-$. Where $H_{a_i}^+ = \{h_1, \ h_2\}, H_{a_i}^- = \{h_3, h_4\}$, with $h_1 < h_2$ and $h_3 > h_4$. Select the fuzzy measure for the generating term and hedge.
(4)     $D_{a_i} = [\min_{a_i}, \max_{a_i}]$ // $\min_{a_i}$ , $\max_{a_i}$ : min and max value of domain a$_i$.
(5)     $FD_{a_i} = H_{a_i}(c_{a_i}^+) \cup H_{a_i}(c_{a_i}^-)$ .
(6) **End**

(7) Determine *intervals level k* of fuzzy condition: $k\mathcal{Q}$.

// Partition $D_{a_i}$ into interval *similar level k*.

(8) $k = k\mathcal{Q}$; // level partition largest with k = 4
(9) **For** i = 1 **to** p **do**
(10)     **For** j = 1 **to** $2^5$(k-1) **do**
(11)         Construct intervals similar level k: $S_{a_i}^k(x_j)$ ;

// Determine *neighborhood level k* of *o.a$_i$*.

(12) **For** each o $\in$ C **do**
(13)     **For** i = 1 **to** p **do**
(14)     **Begin**
(15)         t=0;
(16)         **Repeat**
(17)             t=t+1;
(18)         **Until** $o.a_i \in S_{a_i}^k(x_t)$ or t > $2^5$(k - 1);
(19)         $FRNA_i^k(attr_i) = FRNA_i^k(attr_i) \cup S_{a_i}^k(x_t)$ ;
(20)     **End**

// Determine *neighborhood level k* of *fzvalue$_i$*.

(21) **For** i = 1 **to** p **do**
(22) **Begin**
(23)     t=0;
(24)     **Repeat**
(25)         t=t+1;
(26)     **Until** $fzvalue_i \in S_{a_i}^k(x_t)$ or t > $2^5$(k - 1);
(27)     $FRN_i^k(fzvalue_i) = FRN_i^k(fzvalue_i) \cup S_{a_i}^k(x_t)$ ;
(28) **End**
(29) result=$\varnothing$ ;
(30) **For** each o $\in$ C **do**
(31)     **if** $\underset{i=1}{\overset{p}{\vartheta}}$ ( $FRNA(attr)_i^k = FRNA(fzvalue)_i^k$ )
        **then** result=result $\cup$ {o};
(32) Return result;

Similar to the method we have **SMSN** algorithm following:

**SMSN algorithm:** *search data cases single fuzzy conditions for method.*

Search data in this case, the first we determine *neighborhood level k* fuzzy conditions of method is *FRNP$_k$(fzpvalue)*. Further, we determine *neighborhood level k* of attributes which method handing: *FRNA$_k$(attr1), FRNA$_k$(attr2), …, FRNA$_k$(attrn)*. We choose the function combination of hedge algebras being consistent with method that it operate. Then, *neighborhood level k* of function combination is *FRNPA$_k$(x)*.

At last based on definition 3.1, we perform data matching two *neighborhood level k* of *FRNP$_k$(fzpvalue)* and *FRNPA$_k$(x)*.

**Input**: A class C = ({a$_1$, a$_2$, …, a$_n$}, {M$_1$, M$_2$, …, M$_m$}), C = { o$_1$, o$_2$,…, o$_n$}.

where a$_i$, i = 1…p is attribute, M$_j$ is methods.

**Output**: Set of objects o $\in$ C satisfy condition (o.M$_i$ = *fzpvalue* ).

**Method**

// Initialization.

(1) **For** i = 1 **to** p **do**

(2) **Begin**

(3) Set $G_{a_i} = \{\, 0,\ c_{a_i}^-,\ W,\ c_{a_i}^+,\ 1\,\}$; $H_{a_i} = H_{a_i}^+ \cup H_{a_i}^-$.

Where $H_{a_i}^+ = \{h_1, h_2\}$, $H_{a_i}^- = \{h_3, h_4\}$, with $h_1 < h_2$ and $h_3 > h_4$. Select the fuzzy measure for the generating term and hedge.

(4) $D_{a_i} = [\min_{a_i}, \max_{a_i}]$ // $\min_{a_i}$, $\max_{a_i}$: min and max value of domain $a_i$.

(5) $FD_{a_i} = H_{a_i}(c_{a_i}^+) \cup H_{a_i}(c_{a_i}^-)$.

(6) **End**

(7) Determine *intervals level k* of fuzzy condition: $kQ$.

//Partition $D_{a_i}$ into interval *similar level k*.

(8) $k = kQ$; // level partition largest with $k = 4$

(9) **For** i = 1 **to** p **do**

(10) **For** j = 1 **to** $2^5(k-1)$ **do**

(11) Construct intervals similar level k: $S_{a_i}^k(x_j)$ ;

//Determine *neighborhood level k* of $o.a_i$.

(12) **For** each o $\in$ C **do**

(13) **For** i = 1 **to** p **do**

(14) **Begin**

(15) t=0;

(16) **Repeat**

(17) t=t+1;

(18) **Until** $o.a_i \in S_{a_i}^k(x_t)$ or t $> 2^5(k\text{-}1)$;

(19) $FRNA_i^k(attr_i) = FRNA_i^k(attr_i) \cup S_{a_i}^k(x_t)$ ;

(20) **End**

// Determine *neighborhood level k* of *fzpvalue*.

(21) $i = 1;\ f = 0;$

(22) **While** (i<=p) and (f = 0) **do**

(23) **Begin**

(24) j=0;

(25) **While** (j<=$2^5(k-1)$) and (f = 0) **do**

(26) **Begin**

(27) j=j+1;

(28) **if** *fzpvalue* $\in S_{a_i}^k(x_j)$ **then** f = 1;

(29) **End**;

(30) i = i + 1;

(31) **End**

(32) $FRNP^k(fzpvalue) = S_{a_i}^k(x_j)$ ;

(33) **For** each o $\in$ C **do**

(34) **For** i=1 **to** m **do**

(35) function combination hedge algebras:

$FRNPA_i^k(x_i) = \overset{p}{\underset{j=1}{\vartheta}}(FRNA_j^k(attr_i))$ ;

(36) result=$\varnothing$ ;

//Combination of hedge algebras with operation $\vartheta$ is operation *and*

(37) **For** each o $\in$ C **do**

(38) **For** i=1 **to** m **do**

(39) **if** $FRNPA_i^k(x_i) = FRNP^k(fzpvalue)$

(40) **then** result = result $\cup$ {o};

(40) Return result;

**Theorem: SASN** algorithm and **SMSN** algorithm always stop and correct.

*Proof*:

1. *The Stationarity*: Set of attributes, the method of the object is finite (n, p, m is finite) so algorithm will stop when all objects completed the approved.

2. *The corrective maintenance*:

Really, for each attribute $a_i$ ($1 \le i \le n$) in object o $\in$ C, the attribute values can get a classic value (precise value) or linguistic value (fuzzy value). In relation matching for data, we are divided into the following two cases:

*First case*: For classic attribute values (precise value), we use operation = to perform data matching.

*Second case*: For linguistic value, we use operation matching at level $=_k$, with $k$ is interval *neighborhood level k* by hedge algebra. Based on quantitative semantics, we determined *neighborhood level k* of term $x$ is $FRN_k(x)$ = [a, b], the following cases:

a) If y is classic value (precise value) that y $\in$ [a, b] then y $=_k$ x.

b) If y is linguistic value in interval [$x_1$, $x_2$] (it is calculated through quantitative semantics) that a <= $x_1$ and $x_2$ <= b then y $=_k$ x.

Two algorithms are implemented to matching data in case data is classical or linguistic values and the output is corrective.

Computational complexity of **SASN** algorithm evaluation follows as: step (1) - (19) complexity is $O(p)$, step (20) - (32) is $O(n*p)$. So, the **SASN** algorithm can computational complexity $O(n*p)$.

Computational complexity of **SMSN** algorithm evaluation follows as: step (1) - (23) complexity is $O(p)$, step (24) - (32) is $O(n*p)$, step (33) - (36) is $O(m*n*p)$, step (37) - (40) is $O(m*n)$. So, the **SMSN** algorithm can computational complexity $O(n*p*m)$.

## IV. EXAMPLE

We consider a database with six rectangular objects as follows:

| Rectangular | | | |
|---|---|---|---|
| iD | name | length of edges | width of edges | area() |
| iD1 | hcn1 | 62 | `Little short` | |
| iD2 | hcn2 | 53 | 55.5 | |
| iD3 | hcn3 | `very very short` | 70 | |
| iD4 | hcn4 | 58 | `very long` | |
| iD5 | hcn5 | `little long` | 45 | |
| iD6 | hcn6 | 55 | `Little short` | |

**Query 1:** List of rectangles have length "`Little long`" or "`Little short`"

Using algorithms **SASN** the following:

**Step (1) - (6):**

Let consider a linear hedge algebra of *length*, $\underline{X}_{length}$ = ( $\underline{X}_{length}$ $G_{length}$ $H_{length}$ $\le$), where $G_{length}$ = {`short, long`}, $H^+_{length}$ = {`More, Very`}, $H_{length}$ = {`Possibly`,

*Little]*, where *P*, *L*, *M* and *V* stand for *Possibly*, *Little*, *More* and *Very*, with *Very > More and Little > Possibly*.

Suppose that $W_{length}$ = 0.6, *fm(short)* = 0.6, *fm(long)* = 0.4, *fm(V)* = 0.35, *fm(M)* = 0.25, *fm(P)* = 0.2, *fm(L)* = 0.2.

Dom(DODAI) = [0, 100]. Result=$\varnothing$; $LD_{length} = H_{length}(short) \cup H_{length}(long)$.

**Step (7) - (20):** so *little long* and *little short* = 2 so we only need to build interval *similar level 2*. We perform partition the interval [0, 100] into interval *similar level 2*:

*fm(VVshort)* = 0.35 * 0.35 * 0.6 * 100 = 7.35, so *S(0)* = [0, 7.35];

*fm(MVshort) + fm(PVshort)* = (0.25 * 0.35 * 0.6 + 0.2 * 0.35 * 0.6) * 100 = 9.45, so *S(Vshort)* = (7.35, 16.8];

*fm(LVshort) + fm(VMshort)* = (0.2 * 0.35 * 0.6 + 0.35 * 0.25 * 0.6) * 100 = 9.45; *fm(MMshort) + fm(PMshort)* = (0.25 * 0.25 * 0.6 + 0.2 * 0.25 * 0.6) * 100 = 6.75, so *S(Mshort)* = (26.25, 33];

*fm(LMshort) + fm(VPshort)* = (0.2 * 0.25 * 0.6 + 0.35 * 0.2 * 0.6) * 100 = 7.2; *fm(MPshort) + fm(PPshort)* = (0.25 * 0.2 * 0.6 + 0.2 * 0.2 * 0.6) * 100 = 5.4, so *S(Pshort)* = (40.2, 45.6];

*fm(LPshort) + fm(VLshort)* = (0.2 * 0.2 * 0.6 + 0.35 * 0.2 * 0.6) * 100 = 6.6; *fm(MLshort) + fm(PLshort)* = (0.25 * 0.2 * 0.6 + 0.2 * 0.2 * 0.6) * 100 = 5.4, so *S(Lshort)* = (52.2, 57.6];

with similar calculations, we have *S(W)* = (57.6, 61.6]; *S(Llong)* = (61.6, 65.2]; *S(Plong)* = (69.6, 73.2]; *S(Mlong)* = (78, 82.5]; *S(Vlong)* = (88.8, 95.1]; *S(1)* = (95.1, 100];

**Step (21) - (28):** Determine the *neighborhood level 2* of *Little Long* and *Little Short*. We have *Little Long* ∈ *S(Little Long)* so *neighborhood level 2 of Little Long is* $FRN_2(Little\ Long)$ = *S(Little Long)* = (61.6, 65.2], and *neighborhood level 2 of Little Short* is $FRN_2(Little\ Short)$ = *S(Little Short)* = (52.2, 57.6].

**Step (29) - (32):** According to conditions:

- The length *Little Long* so we have two objects satisfied is iD1, iD5.
- The width *Little Short* so we have three objects satisfied is iD1, iD2, iD6.

So result = {iD1, iD2, iD5, iD6} satisfied a query with the operation *or*.

**Query 2:** List of rectangles have area is "*less small*".

Using algorithms **SMSN** the following:

**Step (1) - (6):**

Let consider a linear hedge algebra of *length*, $\underline{X}_{length}$ = ( $\underline{X}_{length}$, $G_{length}$, $H_{length}$, ≤), where $G_{length}$ = *{Short, Long}*, $H^+_{length}$ = *{More, Very}*, $H^-_{length}$ = *{Possibly, Little}*, where *P*, *L*, *M* and *V* stand for *Possibly,*

*Little*, *More* and *Very*, with *Very > More and Little > Possibly*.

Suppose that $W_{length}$ = 0.6, *fm(short)* = 0.6, *fm(long)* = 0.4, *fm(V)* = 0.35, *fm(M)* = 0.25, *fm(P)* = 0.2, *fm(L)* = 0.2.

Dom(DODAI) = [0, 100]. Result=$\varnothing$; $LD_{length} = H_{length}(short) \cup H_{length}(long)$.

**Step (7) - (20):** so *less small* we see it corresponds to *Little Short*, that *Little Short* = 2 so we only need to build interval *similar level 2*. We perform partition the interval [0, 100] into interval *similar level 2*: (similar calculation in query 1)

*S(0)* = [0, 7.35]; *S(VShort)* = (7.35, 16.8]; *S(MShort)* = (26.25, 33]; *S(PShort)* = (40.2, 45.6]; *S(LShort)* = (52.2, 57.6]; *S(W)* = (57.6, 61.6]; *S(LLong)* = (61.6, 65.2]; *S(PLong)* = (69.6, 73.2]; *S(MLong)* = (78, 82.5]; *S(VLong)* = (88.8, 95.1]; *S(1)* = (95.1, 100];

**Step (21) - (32):** Determine the *neighborhood level 2* of *less small*. So *less small* = *Little Short* ∈ *S(Little Short)* so *neighborhood level 2 of less small* is $FRNP_2(Little\ Short)$ = *S(Little Short)* = (52.2, 57.6].

**Step (33) - (40):** According to conditions:

- The length *Little Short* so we have two objects satisfied is iD2, iD6.
- The width *Little Short* so we have three objects satisfied is iD1, iD2, iD6.

The function combined hedge algebra is product of hedge algebra with the operation *and*, so result = {iD2, iD6} satisfied the conditions of query 2.

## V. CONCLUSION

In this paper, we propose a new method for linguistic data processing in object-oriented database that its information is fuzzy and uncertainty approach to the sematic neighborhood based on hedge algebras. This approach makes easy to process data and homogeneous data. Based on quantitative semantics, we determined *neighborhood level k* of linguistic values and perform data matching by *neighborhood level k* this. This paper has proposed a method combination of hedge algebras in case the attribute value is the linguistic value. From data matching based sematic neighborhood of hedge algebras, this paper has proposed two algorithms **SASN** and **SMSN** for searching data with fuzzy conditions based sematic neighborhood of hedge algebras.

AUTHORS PROFILE

Name: Doan Van Thang

Birth date: 1976.

Graduation at Hue University of Sciences – Hue University, year 2000. Received a master's degree in 2005 at Hue University of Sciences – Hue University. Currently a PhD student at Instiute of Information Technology, Academy Science and Technology of Viet Nam.

Research: Object-oriented database, fuzzy Object-oriented database. Hedge Algebras.

Email:vanthangdn@gmail.com